\documentclass[aps,prl,twocolumn,showpacs,amsmath,amssymb,superscriptaddress,floatfix,tighten,a4paper]{revtex4}
\usepackage{graphicx}
\usepackage{times}

\begin{document}

\title{\large Repulsion Between Inorganic Particles Inserted Within Surfactant Bilayers}

\author{Doru Constantin}
\email[Corresponding author. Tel.:~+33 1 69 15 53 94 ; fax:~+33 1 69
15 60 86; e-mail address:~]{constantin@lps.u-psud.fr}
\affiliation{Laboratoire de Physique des Solides, Universit\'{e}
Paris-Sud, CNRS, UMR8502, 91405 Orsay Cedex, France.}

\author{Brigitte Pansu}
\author{Marianne Imp\'{e}ror}
\author{Patrick Davidson}
\affiliation{Laboratoire de Physique des Solides, Universit\'{e}
Paris-Sud, CNRS, UMR8502, 91405 Orsay Cedex, France.}

\author{Fran\c{c}ois Ribot}
\affiliation{Laboratoire de Chimie de la Mati\`{e}re Condens\'{e}e,
Universit\'{e} Paris VI, CNRS, UMR7574, 75252 Paris Cedex 05,
France.}

\date{\today}

\begin{abstract}
We study by synchrotron small-angle X-ray scattering highly aligned
lamellar phases of a zwitterionic surfactant, doped with
monodisperse and spherical hydrophobic inorganic particles as a
function of particle concentration. Analysis of the structure factor
of the two-dimensional fluid formed by the particles in the plane of
the bilayer gives access to their membrane-mediated interaction,
which is repulsive, with a contact value of about 4~$k_B T$ and a
range of 14~{\AA}. Systematic application of this technique should
lead to a better understanding of the interaction between membrane
inclusions.
\end{abstract}


%

\maketitle


In the last decades, much effort was dedicated to the understanding
of self-assembled membranes, and in particular of the interaction
between membrane inclusions and the host bilayer. This is a
challenging problem, since the membrane must be considered as a
many-particle system, its properties being collectively determined
by the assembly and not by the chemical properties of the individual
molecules \cite{Jensen:2004}. Notwithstanding the complexity of the
system, the concepts developed in soft matter physics should be
operative in this context and even `simplified' models could yield
valuable information. For this reason, a considerable body of work
deals with the theoretical modelling and numerical simulation of
such mixed systems, aiming in particular to determine the
membrane-mediated interaction between inclusions
\cite{Sperotto:2006}. Thorough understanding of these interactions
could make a substantial contribution to topics as diverse as the
formulation of new composite systems and understanding the activity
of membrane proteins.

However, this theoretical work was not yet matched by enough
experimental results. The first such data was obtained by directly
measuring the radial distribution function of membrane inclusions
using freeze-fracture electron microscopy (FFEM) \cite{ref3}. These
data were compared to liquid state models and could be described by
a hard-core model with, in some cases, an additional repulsive or
attractive interaction \cite{ref4}. FFEM was not extensively used,
undoubtedly due to the inherent experimental difficulties; moreover,
it is not obvious that the distribution measured in the frozen
sample is identical to that at thermal equilibrium.

Considering the typical length scales to be probed, X-ray and
neutron scattering techniques \cite{ref5} are uniquely adapted to
the study of this problem. As an example, the pores formed by the
antimicrobial peptide alamethicin in dimyristoylphosphatidylcholine
bilayers were shown to repel each other \cite{Constantin:2007}.
These studies are however hindered by the low scattering contrast of
the proteins (and, in many cases, by their scarcity), whence the
practical (and also the fundamental) interest of finding out whether
other --perhaps more adapted-- particles can be inserted within
membranes.

The purpose of this Letter is to show that self-assembled bilayers
can be doped with significant amounts of (hydrophobic and
charge-neutral) hybrid nano-objects and that these probes can be
used to determine accurately the membrane-mediated interaction.
Their use presents substantial advantages: they are ``rigid'' (fixed
atomic configuration) and perfectly monodisperse, therefore imposing
a well-defined membrane deformation, whereas membrane proteins can
assume various conformations; their scattering contrast is high (due
to the presence of metal atoms); their surface properties can be
tailored by changing the nature of the grafted ligands.
Conceptually, these inclusions are also easier to model, since they
do not ``break'' the surface of the monolayer and hence there is no
contact line, where the choice of the boundary condition would be
somewhat delicate \cite{Harroun:1999b}.

This approach can help clarify long-standing questions, such as: Is
a continuous model sufficient for an accurate description of the
membrane and, if so, down to what length scale? What are the
relevant parameters and how can they be measured? What are the
specificities of mixed bilayers, and in particular of lipid
membranes (microscale separation, raft formation etc.)? In the long
run, systematic studies should yield a clearer picture of membranes
as two-dimensional complex systems.


The nanoparticles used here are butyltin oxo clusters
\{(BuSn)$_{12}$O$_{14}$(OH)$_6$\}$^{2+}$(4-CH$_3$C$_6$H$_4$SO$_3^-$)$_2$,
denoted by BuSn12 in the following. They were synthesized and
characterized as described in reference \cite{Eychenne:2000} (for
their structure, see Figure~1 in this reference). They were
dissolved in ethanol at a concentration of 23.47~wt.\%.

The zwitterionic surfactant, dimethyldodecylamine-N-oxide (DDAO) was
purchased from Sigma-Aldrich and dried in vacuum (using a liquid
nitrogen-cooled solvent trap) for 20~h. No weight loss was observed
after this step, so we conclude that the surfactant was dry as
supplied (see \cite{Kocherbitov:2006} for a detailed discussion).
DDAO was then dissolved in isopropanol at a concentration of
23.02~wt.\%.

We mixed the BuSn12 and DDAO solutions to yield the desired particle
concentrations, and then dried the mixtures in vacuum; the final
(dry) mass was about 200~mg for each sample. We then added water at
a concentration of 15--27~wt.\% of the final hydrated mixtures,
which are thus in the fluid lamellar $L_{\alpha}$ phase (see the
phase diagram of the undoped system in \cite{Kocherbitov:2006}). The
molecular weight of DDAO is 229.40 (Sigma-Aldrich), its density is
$0.84 \mathrm{g/cm}^3$ and the thickness of the bilayer is $25 \pm 1
\mathrm{\AA}$ \cite{ref10}, yielding an area per surfactant molecule
$A_{\mathrm{DDAO}} = 37.8 \mathrm{\AA}^2$. For the BuSn12 particles,
we take a molecular weight of 2866.7 and a density of $1.93
\mathrm{g/cm}^3$ \cite{Eychenne:2000}. Using these values and
neglecting the increase in bilayer surface due to the inserted
particles yields the (two-dimensional) number density of particles
in the plane of the membrane, $n$.

The samples were prepared in flat glass capillaries (VitroCom Inc.,
Mt. Lks, N.J.), 100~$\mu$m thick and 2~mm wide by gently sucking in
the lamellar phase using a syringe. The capillaries were
flame-sealed. Good homeotropic alignment (lamellae parallel to the
flat faces of the capillary) was obtained by thermal treatment,
using a Mettler~FP52 heating stage. The samples were heated into the
isotropic phase (at $130 {\,}^{\circ} \mbox{C}$) and then cooled
down to the lamellar phase at a rate of $1 {\,}^{\circ} \mbox{C} /
\mathrm{min}$.

The small-angle x-ray scattering measurements were performed at the
European Synchrotron Radiation Facility (ESRF, Grenoble, France) on
the bending magnet beamline BM02 (D2AM), at a photon energy of
11~keV. See reference \cite{Simon:1997} for more details on the
setup. The data was acquired using a CCD Peltier-cooled camera
(SCX90-1300, from Princeton Instruments Inc., New Jersey, USA) with
a resolution of $1340 \times 1300$ pixels. Data preprocessing (dark
current subtraction, flat field correction, radial regrouping and
normalization) was performed using the \texttt{bm2img} software
developed at the beamline.

The incident beam was perpendicular to the flat face of the
capillary (parallel to the smectic director, which we take along the
$z$ axis.) Thus, the scattering vector $\mathbf{q}$ is mostly
contained in the $(x,y)$ plane of the layers, and the measured
scattered signal $I(\mathbf{q})$ probes inhomogeneities of the
electron density in this plane. Since the bilayers form a
two-dimensional liquid, the scattering pattern exhibits azimuthal
symmetry: $I=I(q=|\mathbf{q}|)$. We also measured the scattering
intensity of two BuSn12/ethanol solutions in the same type of
capillary. The accessible scattering range was $0.04 < q < 0.9
\mathrm{\AA}^{-1}$.
\begin{figure}
\includegraphics[width=8 cm,angle=0]{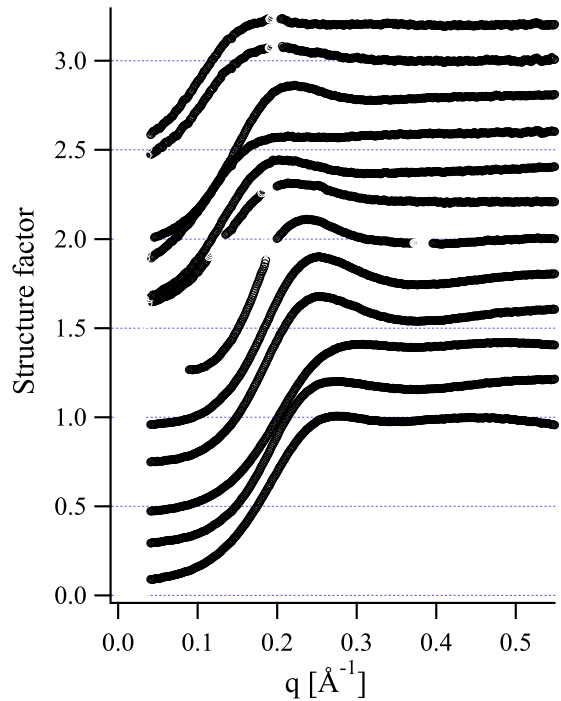}
\caption{Structure factors $S(q)$ of the 2D fluid formed by the
BuSn12 particles in the plane of the membranes, for different
concentrations, from top to bottom: $n=0.217$, 0.231, 0.429, 0.434,
0.451, 0.584, 0.801, 1.102, 1.275, 1.976, 2.147, and $2.304 \,
10^{-3} \mathrm{\AA}^{-2}$. The curves are shifted vertically in
steps of 0.2. Gaps in the curves correspond to the presence of
(weak) lamellar peaks, due to occasional alignment defects, mainly
taking the form of oily streaks.} \label{fig:structure}
\end{figure}
Since the electron density of the butyl chains is similar to that of
the dodecyl chains within the bilayers and to that of ethanol, we
expect the electron contrast of the particles to be due exclusively
to their inorganic core, which is slightly ovoidal, with an average
radius of 4.5~\AA. Indeed, the intensity at higher scattering
vectors ($q
> 0.5 \mathrm{\AA}^{-1}$) is well described for all samples by the
form factor of a sphere $|Ff(R,q)|^2$, with a radius $R = 4.5 \pm
0.2 \mathrm{\AA}$ used as a free fitting parameter. The interaction
between particles is described by the structure factor, defined as
$S(q) = I(q) / |Ff(R,q)|^2$ \cite{Chaikin95}. The structure factors
thus obtained are shown in Figure \ref{fig:structure} for all
in-plane concentrations, listed in the caption.


The additional interaction is viewed as a perturbation with respect
to the hard core (disk or sphere) model, taken into account via the
random phase approximation (RPA) \cite{Andersen:1970}. In this
approach, one obtains the direct correlation function of the
perturbed system $c(r)$ from that of the reference system
$c_{\mathrm{ref}}(r)$ as: $c(r) = c_{\mathrm{ref}}(r) - \beta U(r)$
\cite{Hansen:1986} or, equivalently:
\begin{equation} n \beta \widetilde{U}
(q) = S^{-1}(q) - S_{\mathrm{ref}}^{-1}(q)
\label{eq:RPA}\end{equation} with $\beta=(k_B T)^{-1}$.

In three dimensions, the reference structure factor
$S^{3D}_{\mathrm{ref}}$ for BuSn12 particles dissolved in ethanol is
given by a hard sphere interaction (in the Percus-Yevick
approximation \cite{Wertheim63,Thiele63}) with a hard-core radius of
4.5~{\AA}; the numerical particle density $n_{3D}$ (in 3D) is
determined from the mass concentration of the solutions.

In two dimensions, an analytical form for the structure factor
$S^{2D}_{\mathrm{ref}}$ of hard disks was given by Rosenfeld
\cite{Rosenfeld90}; we use the same core radius of 4.5~{\AA} as
above. The Fourier transform of the interaction potential,
$\widetilde{U} (q)$, obtained by applying relation (\ref{eq:RPA}) to
the data in Fig. \ref{fig:structure}, is shown in Fig. \ref{fig:Vq}
for all concentrations. For ease of calculation, both in two and
three dimensions we model the interaction $U(r)$ by a Gaussian, with
amplitude $U_0$ and range $\xi$:
\begin{figure}
\includegraphics[width=8 cm,angle=0]{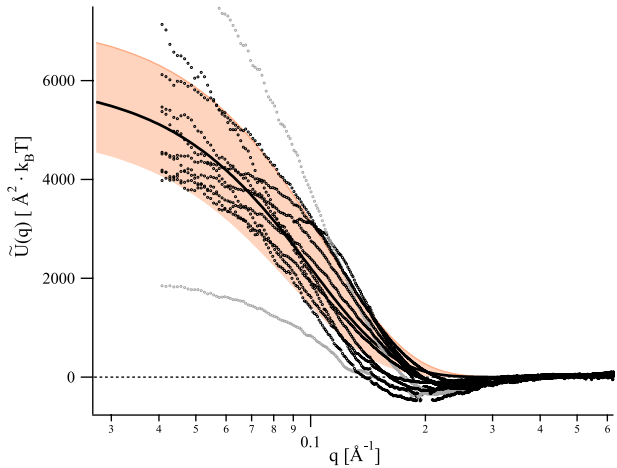}
\caption{The Fourier transform of the interaction potential,
$\widetilde{U} (q)$, (open dots) obtained from the structure factors
in Figure \ref{fig:structure} in the RPA approximation
(\ref{eq:RPA}). The solid line is a Gaussian model. The shaded area
is chosen to cover most experimental points and is also delimited by
two Gaussians (see text for details). The data points far above and
far below the shaded range correspond to $n=0.429$ and $0.584 \,
10^{-3} \mathrm{\AA}^{-2}$, respectively (third and sixth curve from
the top in Figure \ref{fig:structure}).} \label{fig:Vq}
\end{figure}
\begin{equation} U(r) = U_0 \exp \left [ - (r / \xi)^2 \! / 2 \right ] \label{eq:ur} \end{equation}
\noindent with $\widetilde{U} (q)$ the Fourier transform of $U(r)$,
given by:
\begin{equation}
\label{eq:uq} \widetilde{U} (q) =
\begin{cases}
2 \pi \, U_0 \, {\xi}^2 \exp \left [- \left (
q \xi \right )^2 \! /2 \right ]  & \mathrm{in~2D}\\
(2 \pi)^{3/2} \, U_0 \, {\xi}^3 \exp \left [- \left ( q \xi \right
)^2 \! /2 \right ]  & \mathrm{in~3D}
\end{cases}
\end{equation}
With the exception of two curves (for $n=0.429$ and $0.584 \,
10^{-3} \mathrm{\AA}^{-2}$) that are anomalously high or low,
respectively, all data points are reasonably well covered by the
shaded area in Figure \ref{fig:Vq}, limited by two Gaussians (2D
case in Eq. \ref{eq:uq}), with $U_0 = 3.22 \, k_B T$ and $\xi =
15.71 \, \mathrm{\AA}$ for the lower bound and $U_0 = 6.93 \, k_B
T$, $\xi = 12.86 \, \mathrm{\AA}$ for the upper bound. The solid
line, giving approximately the midline of the shaded area, is
described by $U_0 = 4.75 \, k_B T$ and $\xi = 14.14 \,
\mathrm{\AA}$. The apparent negative values of $\widetilde{U} (q)$
around $q=0.2 \, \mathrm{\AA}^{-1}$ in Figure \ref{fig:Vq} are
probably due to the enhancement of the interaction peak of $S(q)$
with respect to the reference potential by the presence of the
repulsive interaction; this feature is not captured by the simple
RPA treatment.
\begin{figure}
\includegraphics[width=8 cm,angle=0]{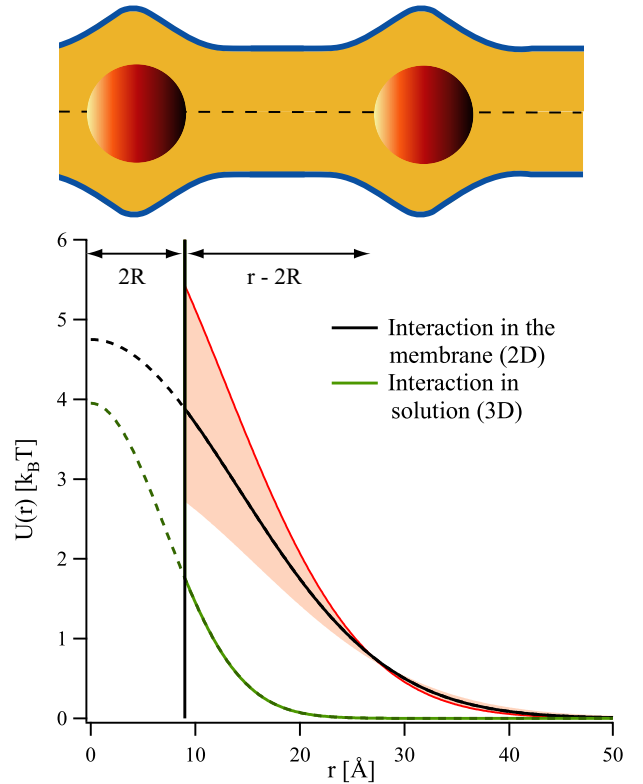}
\caption{Interaction potential $U(r)$ of the BuSn12 particles within
the bilayers, obtained by taking the inverse Fourier transform of
the solid line and the shaded area in Figure \ref{fig:Vq} (see text
for the numerical values). The lower curve is the interaction
potential of the particles in ethanol. The solid vertical line marks
the hard core interaction with radius 4.5~{\AA}}
\label{fig:potential}
\end{figure}
The (real-space) interaction potentials $U(r)$ corresponding to the
aforementioned values of $U_0$ and $\xi$ are plotted in Figure
\ref{fig:potential}, using the same convention. Thus, the solid line
is the estimate for the interaction and the shaded area the
uncertainty due to the value spread in Figure \ref{fig:Vq}. Of
course, for $r < 2R=9 \, \mathrm{\AA}$ the repulsion is due to the
hard core (solid vertical line).

To make sure that the interaction is induced by the membrane, we
also show the repulsion measured for BuSn12 particles in ethanol
(lower solid line in Figure \ref{fig:potential}); its amplitude and
range are clearly lower than for the interaction in the membrane
(for reference, $U_0 = 4 \, k_B T$ and $\xi = 7 \, \mathrm{\AA}$).
It is probably due to the steric repulsion between the butyl chains
grafted onto the inorganic core.

The main result of this work is hence that the (two-dimensional)
interaction between BuSn12 particles inserted within DDAO bilayers
can be described by a potential of the form (\ref{eq:ur}), with $U_0
= 5 \pm 1 \, k_B T$ and $\xi = 14 \pm 1 \, \mathrm{\AA}$. Below, we
discuss briefly the various theoretical predictions, but a direct
comparison cannot be made since they generally concern cylindrical
inclusions that traverse the membrane, without being covered by the
monolayers.


The simplest approach is to model the membrane as a continuous
medium and write the free energy of the bilayer+inclusions system in
terms of elastic deformation, described by the Helfrich Hamiltonian
\cite{Helfrich:1973}, where the molecular properties of the bilayer
are abstracted into mesoscopic parameters, such as the bending and
stretching moduli and the spontaneous curvature of the monolayer. A
systematic study was performed in the group of Pincus
\cite{Dan:1993,Dan:1994,Aranda:1996}. Their results highlight the
importance of the spontaneous curvature of the monolayers $c_0$: the
interaction is attractive when $c_0$ vanishes (the elastic energy is
minimized by the aggregation of inclusions) but can become repulsive
when the deformation induced by the inclusion is such that the
curvature of the individual monolayer has the sign of $c_0$. In our
case, this would mean that the DDAO monolayers tend to have $c_0 >
0$ (convex head-groups, see the sketch in Figure
\ref{fig:potential})

However, inclusions also perturb the structure of the membrane by
restricting the conformation of the lipid chains in their vicinity
\cite{ref22}, and this can lead to significant interaction, even in
the absence of ``large scale'' bilayer deformation (no hydrophobic
mismatch.) In particular, Lag{\"{u}}e et al. \cite{ref23} find that
``smooth'' hard cylinders repel each other in some lipid bilayers;
the amplitude and range of this interaction is in qualitative
agreement with our results (for cylinders with a 5~{\AA} radius in
dioleoylphosphatidylcholine bilayers, the repulsive lipid-mediated
interaction has a maximum value of $7 \, k_B T$ and extends 20~{\AA}
from contact.) Nonetheless, the interaction is highly dependent on
the chemical structure of the lipid, so a meaningful comparison is
difficult to make using the available data.

Recently, some efforts were made \cite{May:2002,Bohinc:2003} to
account for both effects (elastic energy and restrictions on chain
conformations) within an extended model, including as variables both
the variation in membrane thickness and the local molecular tilt.
Their coupling removes the symmetry between the positive and
negative spontaneous curvature values, leading to repulsion only for
$c_0 > 0$ combined with negative hydrophobic mismatch (bilayer
``pinching''.) At first sight this prediction seems to be at odds
with our system, but the difference in geometry plays an important
role: in our case, we expect the particles to be covered by
positively curved monolayer caps (energetically favourable), while
the calculation in \cite{Bohinc:2003} considers the inclusions as
vertical cylinders. A more accurate comparison should be very
interesting.


We have shown that nano-objects can be used to probe the properties
of self-assembled bilayers; conversely, one can envision using
ordered surfactant phases (combining a high degree of order with
easy processability and very good wetting properties) to organize
and align such objects in view of applications.

\begin{acknowledgments}
The ESRF is gratefully acknowledged for the award of beam time
(experiment 02-01-732) and we thank C. Rochas for competent and
enthusiastic support. A. Dessombz and A. Poulos are acknowledged for
helping with the synchrotron experiments.

\end{acknowledgments}

\end{document}